\documentclass[pre,twocolumn,preprintnumbers,amsmath,amssymb,floatfix]{revtex4}
\usepackage{graphicx}
\usepackage{bm}

\begin{document}
\title{Colloidal particles at a nematic-isotropic interface: effects of confinement}

\author{John L. West}
\affiliation{Liquid Crystal Institute, Kent State University, Kent, Ohio 44242, USA}

\author{Ke Zhang}
\affiliation{Liquid Crystal Institute, Kent State University, Kent, Ohio 44242, USA}

\author{Anatoliy Glushchenko}
\affiliation{University of Colorado at Colorado Springs, Colorado Springs, CO 80933, USA}

\author{Denis Andrienko}
\affiliation{Max-Plank Institute for Polymer Research, Ackermannweg 10, Mainz 55128, Germany}
\affiliation{Liquid Crystal Institute, Kent State University, Kent, Ohio 44242, USA}

\author{Mykola Tasinkevych}
\affiliation{Max-Planck-Institut f\"{u}r Metallforschung, Heisenbergstr. 3, Stuttgart 70569, Germany}
\affiliation{Institut f\"{u}r Theoretische und Angewandte Physik,
         Universit\"{a}t Stuttgart, Pfaffenwaldring 57, Stuttgart 70569, Germany}

\author{Yuri Reznikov}
\affiliation{Institute of Physics, Prospect Nauky 46, Kyiv 03039, Ukraine}

\date{\today}

\begin{abstract}
  When captured by a flat nematic-isotropic interface, colloidal
  particles can be dragged by it. As a result spatially periodic
  structures may appear, with the period depending on a particle mass,
  size, and interface velocity~\cite{west.jl:2002}.  If liquid crystal
  is sandwiched between two substrates, the interface takes a
  wedge-like shape, accommodating the interface-substrate contact
  angle and minimizing the director distortions on its nematic side.
  Correspondingly, particles move along complex trajectories: they are
  first captured by the interface and then `glide' towards its vertex
  point.  Our experiments quantify this scenario, and
  numerical minimization of the Landau-de Gennes free energy allow for
  a qualitative description of the interfacial structure and the drag
  force.
%  We propose that drag on the particles can help to manipulate their
%  spatial distribution in the cell.
%   as well as to influence the
%  process of flocculation of colloidal particles upon cell cooling.
\end{abstract}

\pacs{61.30.Jf, 64.70.Md, 82.70.2y}

\maketitle

\section{Introduction}
Precise manipulation of tiny particles in liquids is a rapidly
developing direction of a modern technology. Manufacturing of $e$-papers
and electrophoretic displays~\cite{zehner.r:1951,ogawa.m:1951},
separation of bacterial species and living
cells~\cite{pohl.ha:1951,pohl.ha:1978,markx.gh:1996}, trapping of DNA
and polymer
particles~\cite{washizu.m:1990a,washizu.m:1990b,morgan.h:1999}, growth
of photonic crystals~\cite{yoshinaga.k:1999,kralchevsky.p:1994}, are
only a few examples revealing its scientific and technological
importance.

In order to move, organize, or separate colloidal particles several
techniques have been suggested, among them the drag of the micro-particles
by a moving nematic-isotropic interface~\cite{west.jl:2002}. Briefly,
due to the interfacial and surface tensions as well as long-range
distortions of the director field, particles can be captured and
subsequently dragged by the moving interface. Matching the speed of
the interface, particle size, and elastic properties of the liquid
crystal, one can move particles of specified radius and control their
spatial distribution in the cell.

In our previous work~\cite{west.jl:2002} we consider the situation
when the interface between the nematic and isotropic phases is
flat. In a slab geometry, however, the interface bends
accommodating the contact angle between the nematic phase, the isotropic
phase, and the substrates, and minimizing the director distortions on
the nematic side of the interface~\cite{oswald.p:1987,MullolBLO99}. As
a result, the meniscus of the isotropic phase extends into the nematic
phase; this of course influences particle trajectories in the
interfacial region.
\begin{figure}
\begin{center}
\centerline{\includegraphics[width=8cm]{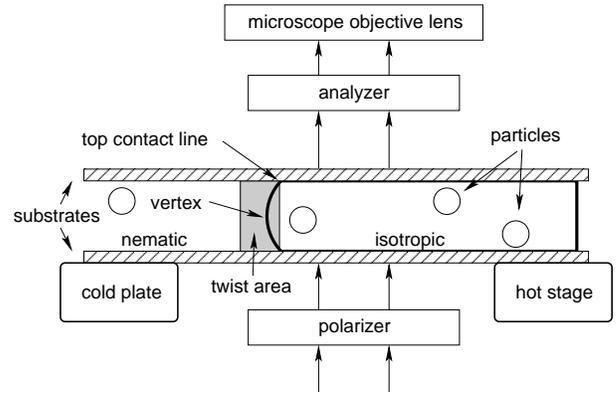}}
\end{center}
\caption{Experimental setup.}
\label{fig:setup}
\end{figure}
In this paper we study the behavior of colloidal particles captured
by a wedge-shaped interface. We first examine the shape of the
interface and the director distribution in its vicinity.
Then we analyze how these factors influence the trajectory of a particle in the
interfacial region. We conclude that there is an additional force on
a particle acting towards the vertex of the interface.
Finally, we compare our experimental results to the estimates based on the
minimization of the Landau-de Gennes free energy.

%%%%%%%%%%%%%%%%%%%%%%%%%%%%%%%%%%%%%%%%%%%%%%%%%%%%%%%%%%%%%%%%%%%%%%%
\section{Experiment}

Spherical polystyrene particles of diameter $d = 16\, \mu \rm m$ are
dispersed in the isotropic phase of liquid crystal 5CB using
ultrasonic shaker. The density of the particles, $1.04\, \rm g /
cm^3$, is slightly higher than the density of pure 5CB, $1.007\, \rm
g/ cm^3$~\cite{blinov.lm:1994}. The surface of the particles provides
planar alignment of the liquid crystal director. $40\, \mu \rm
m$-thick cells are filled with the mixture in the isotropic phase and
cooled to room temperature. The uniform planar alignment is obtained
by rubbing layers of polyamide PI2555 deposited on the substrates.

The experimental setup is sketched in Fig.~\ref{fig:setup} and is
described in detail in Ref.~\cite{west.jl:2002}.  Briefly, one end of
the cell is placed on a hot stage; the other end is attached to a
metal plate, kept at a room temperature.  The temperature of the hot
plate could be changed with the speed $0.1\, \rm deg/min$ and is
monitored with the accuracy $0.05\, \rm deg$.  Typical temperature
gradients over the cell are $20-30\, \rm deg/cm$.  Heating/cooling of
the hot stage with the speed $5\, \rm deg / min$ results in the
interface movement with the speed $v \approx 13\, \mu \rm m / s$. The
system is observed through the polarizing microscope.

\begin{figure}
\begin{center}
\centerline{\includegraphics[width=8cm]{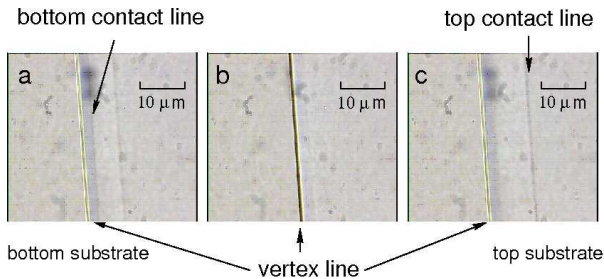}}
\end{center}
\caption{
Typical optical images of the extended interface. The microscope
objective lens is focused on a) bottom substrate; b) vertex line, cell
bulk; c) top substrate. }
\label{fig:images}
\end{figure}
Typical optical images of the cell are shown in Fig.~\ref{fig:images}.
Two homogeneous phases are separated by a rather wide, birefringent,
region. The width of this region depends on the speed of the
interface.  In a stationary case it is about $10\, \mu \rm m$, which
corresponds to $0.025^0\, \rm C$; at $13\, \mu \rm m / s$ it is about
$40\, \mu \rm m$.
The region occupied by the extended interface has a distinct boundary
which separates it from the nematic side of the cell. We denote this
boundary as a vertex line.  Focusing the microscope lens on the cell
substrates, we observed two additional lines positioned on the right
side of the vertex line, in the isotropic phase, see
Figs.~\ref{fig:images}(a) and (c), contact lines. The vertex line is
also visible here, however it is less sharp compared to the situation
when the focus is on the bulk of the cell, Fig.~\ref{fig:images}(b).

These observations allow us to reconstruct the shape of the interface.
Since the nematic phase of 5CB wets polyamide-treated
surfaces~\cite{reznikov.ya:1992,wen.b:2002}, the interface bends in
order to accommodate the zero contact angle at the cell
substrates~\cite{oswald.p:1987,bechhoefer.j:1989,wen.b:2002}.
This bending influences the director distribution in the cell.  The
director is anchored at the interface with the tilt angle $\theta
\approx 26.5\, \rm deg$ and rather strong anchoring, $W \approx
10^{-3}\, \rm erg / cm^{2}$~\cite{faetti.s:1984.b}. The substrates
provide planar orientation of the director. To match the boundary
conditions at the interface and the substrates, the director
orientation must change from the uniform planar orientation far from
the interface to the tilted orientation at the interface.

\begin{figure}
\begin{center}
\centerline{\includegraphics[width=8cm]{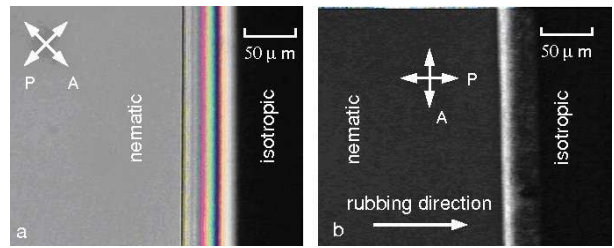}}
\end{center}
\caption{ Snapshot of the moving interface. Hot stage cooling
speed $5 \rm deg/min$.  Orientations of the polarizer (P) and analyzer
 (A) are marked by the white arrows. Rubbing direction is parallel to
 the horizontal side of the pictures. }
\label{fig:twist}
\end{figure}

Rotating the cell between crossed polarizers we have also noticed that
the director is parallel to the rubbing direction everywhere except
close to the vertex line, see Fig.\ref{fig:twist}(a). This region
looked bright between crossed polarizers on the background of the
remaining, dark cell, see Fig.\ref{fig:twist}(b). Rotating the
analyzer we were able to darken the region next to the vertex line.
From these observations we concluded that, close to the vertex line,
the the director twists in the $xz$ plane, parallel to the substrates.
The value of the twist angle was not well reproducible, changing from
$45$ to $90$ degrees.

Director twist can be understood qualitatively, in terms of the
splay-bend and twist contributions to the elastic energy. Next to the
vertex line the director has significant splay-bend deformation, which
is energetically unfavorable. It is known that the director can relax
this deformation by twisting in the perpendicular
plane~\cite{lavrentovich.od:1992,andrienko.d:2004b,stark.h:1999,rudinger.a:1999}.
The balance of the elastic energies of the twist and splay-bend
deformations defines if twist occurs or not. Since the twist elastic
constant $K_{22}$ for 5CB is about two times smaller than the splay
and bend elastic constants, $K_{11}$ and $K_{33}$, it is energetically
more favorable for the director to twist than to bend.
In fact, such an `escape' of the director has already been observed in
a similar system, where the nematic is in the contact with its own
melt~\cite{oswald.p:1987} or with air~\cite{MullolBLO99}.

%We also note that other mechanisms can contribute to the structure of
%the interface. 5CB is a mixture, i.~e. nematic and isotropic phases
%can coexist over a rather wide range of
%temperatures~\cite{palffymuhoray:1982,martire.de:1976,oweimreen.ga:1980,martire.de:1980,reznikov.ya:1992},
%which slows down propagation of the interface in a cell. In fact,
%controlled addition of impurities allows to capture the motion of curved interfaces separating
%bubbles of the nematic phase from the surrounding liquid
%crystal~\cite{vollmer.d:2004}.

\begin{figure}
\begin{center}
\centerline{\includegraphics[width=7cm]{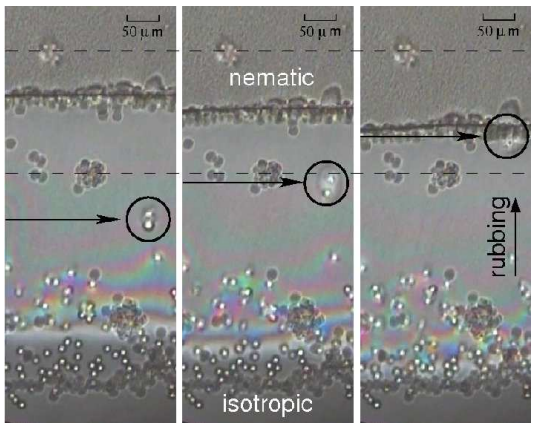}}
\end{center}
\caption{ Snapshots of a moving nematic-isotropic interface, separated
  by the $2 s$ time intervals, which illustrate particle attraction to
  the vertex line before it is captured by the interface. Two dotted
  lines show the position of two aggregates which are stuck to the
  cell surface and serve as reference points.  The interface moves
  slowly from the top to the bottom; at the same time, a cluster of
  two colloidal particles (encircled by a solid line) moves towards
  the interface, from the bottom to the top.  }
\label{fig:attraction}
\end{figure}

%\subsection{Drag on the particles}
Now that the shape of the interface and the director distribution in
the interfacial region are known, let us have a look at the motion of
colloidal particles.  Upon cooling of the hot stage, the area occupied
by the nematic phase grows; the interface moves towards the hot stage,
meeting the particles dispersed in the isotropic phase.  Fast moving
interfaces with the velocity above $18\, \mu \rm m/s$ are not able to
capture the particles. At smaller velocities, dispersed in the
isotropic phase particles are attracted to the interface as soon as
they reach one of the contact lines, as illustrated in
Fig.~\ref{fig:attraction}. Immediately after that they speed up
towards the vertex line, in the direction opposite to the interface
propagation. Upon reaching the vertex line, particles reverse the
direction of motion and start following the interface. 
In fact, even when the particles are deep in the isotropic phase, there
is a small force acting on them in the direction opposite to
the temperature gradient, i.e., towards the interface. 
This force is due to increase of the free
energy of a para-nematic phase at the surface of the particle 
when it moves into the region with a higher temperature.

Upon heating, the interface moves away from the hot stage, pushing the particles
from the isotropic to the nematic phase.  Particles start to interact
with the interface only at a distance comparable to their own diameter
and much smaller speeds of the interface, below $4\, \mu \rm m/s$, are
required to capture them.
%

%As we can see, some of the results confirm our previous
%studies~\cite{west.jl:2002}: particles can be captured by the
%interface, dragged by it, as well as dropped at a certain place. Each
%particular scenario is extremely sensitive to the direction and
%magnitude of the interface velocity, particle size, etc. However,
%there are also a few differences between the drag on the particle by the
%flat and curved interfaces.

\begin{figure}
\begin{center}
\centerline{\includegraphics[width=4cm]{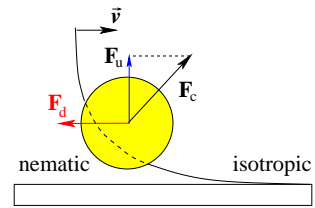}}
\end{center}
\caption{ Schematic representation of the capillary force ${\bf F}_c$
  and the drag force ${\bf F}_d$ exerted on a colloidal particle by
  the moving nematic-isotropic interface. Projection ${\bf F}_u$ of
  the capillary force onto the $y$ axis (perpendicular to the
  substrates) pushes the particle towards the vertex point.}
\label{fig:capillary_force}
\end{figure}

To understand particle dynamics we first note that the density of the
particles is slightly bigger than the density of the liquid crystal.
As a result, they gradually sediment at the bottom substrate, where
they meet the moving interface. If a particle is trapped by the
interface, the interfacial tension results in a force on the particle,
directed perpendicular to the interface. 

Since the particle is moving with the interface it experiences Stokes
drag force in the direction opposite to the velocity of the interface.
If the velocity of the particle does not change, then the sum of all
forces on the particle is zero. This means that the elastic plus
capillary forces shall compensate the drag force. However, the latter
has only the $x$-component, in the direction of the interface
velocity. At the same time, the capillary force is perpendicular to
the interface and has a component pointing towards the vertex point,
as shown in Fig.~\ref{fig:capillary_force}. Hence, the particle will
move until this component of the force vanishes, which is at the
vertex point.

%If one assumes that there is
%a component of the force in the direction opposite to the interface
%propagation, i.e., the capillary force is directed inward the growing
%phase, then due to the wedge-like shape of the interface
%has a local tilt with respect to the propagation
%direction, which changes across the cell. Therefore, 
%there should be a component of the force pushing the particle towards the vertex of the
%meniscus, or upwards, as is schematically drawn in the Fig.~\ref{fig:capillary_force}.
%In principle, the capillary force can  be directed either inward or outward the growing phase, thus
%one can not rule out the possibility that the force pushes the colloid downward, or away from
%the interface vertex.   

%
\begin{figure}
\begin{center}
\centerline{\includegraphics[width=7cm]{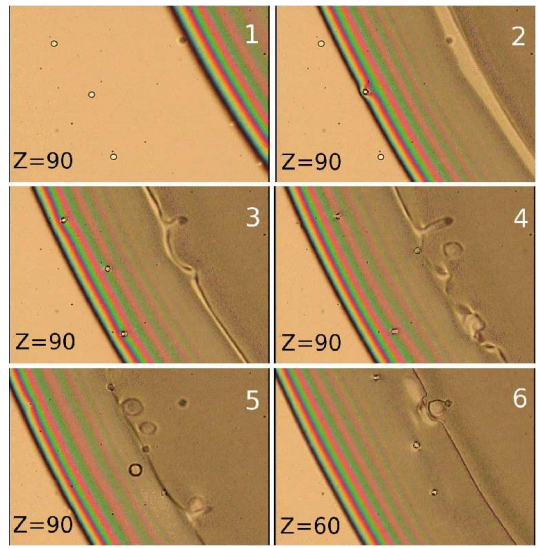}}
\end{center}
\caption{ Snapshots of the moving interface. The three particles are
  attracted to the interface and, because of the vertical motion,
  disappear from the focus (near the NI line, snapshots 4 and 5). In
  order to see them in the focus again, we need to change the focal
  distance from $z=90 \mu m $ to $z=60 \mu m $ (snapshot 6), where $z$ is a vertical
  distance from the objective lens to some arbitrary level.
  Experiments were made with $16 \mu m$
  particles in a $200 \mu m$   thick cell. }
\label{fig:lifting}
\end{figure}

In addition, in order to minimize the elastic free energy of the
system, the particle tries to occupy the place with the strongest
distortions of the director~\cite{voloschenko.d:2002}. This results in
an additional force guiding the particle towards the vertex line.
Together, these two forces lift the particle in the cell, in the
direction opposite to the interface propagation. In fact, the particle
lifting can be directly observed by changing the focal distance of the
objective lens, as shown in Fig.~\ref{fig:lifting}.

To support these arguments, in the following we study the behavior of
the colloid in the vicinity of the confined nematic-isotropic (NI)
interface in the framework of Landau-de Gennes free energy.

\section{Theoretical model and discussion}

\begin{figure}
\begin{center}
\includegraphics[width=6cm, angle=0]{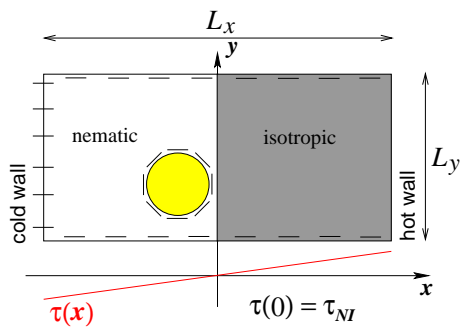}
\end{center}
\caption[Studied geometry]{ Sketch of the $xy$ cross-section of the
  system. System size $L_x\times L_y$. Boundary conditions: planar
  alignment of the nematic director at the particle and cell surfaces.
  Initial conditions: flat nematic-isotropic interface at $x=0$.}
\label{fig:geometry} 
\end{figure}

A geometry mimicking the experimental setup is shown in
Fig.~\ref{fig:geometry}. A uniform temperature gradient is imposed
along the $x$ axis with the `hot' wall at $x = L_x/2$ and the `cold'
wall at $x = -L_x/2$.
%  $L_y$ is the width of the system. 
The director at the `cold', bottom and top walls
 is fixed in the $x$-direction, and the absolute value of the 
order parameter at these walls is fixed to the bulk order 
parameter of the nematic phase at two-phase coexistence. The order
parameter at the `hot' wall is set to zero. 
A colloidal particle, which we take to be a long cylinder of radius
$R$, with the symmetry axis parallel to the $z$ axis, imposes rigid planar anchoring boundary 
conditions at its surface. This results in the formation close to the colloid of two topological defects, known as 
boojums \cite{poulin:98}. 
The defects are aligned in the direction of the liquid crystal alignment, i.e., parallel to
the $x$ axes.

The  system is described by the Landau-de Gennes free
energy~\cite{degennes.pg:1995.a}
\begin{equation}
{\cal F}\{\bm Q\} = \int{(f_b + f_e)dV}
\label{eq:nem_free_en}
\end{equation}
where $f_b$ is the bulk free energy density and $f_e$ is the elastic
free energy density. The minimum of the Landau-de Gennes functional
$F\{\bm Q\}$ gives the equilibrium value of the tensor order parameter
$\bm Q$.
Symmetry arguments yield the bulk free energy
density~\cite{stephen.mj:1974.a,degennes.pg:1995.a}
\begin{equation}
f_b = a(T-T^*) {\rm Tr} {\bm Q^2} -
      b {\rm Tr} {\bm Q^3} +
      c \left[{\rm Tr} {\bm Q^2}\right]^2,
\label{eq:f_b}
\end{equation}
where the positive constants $a,b,c$ are taken temperature
independent. It is convenient to scale out the variables by defining
\begin{eqnarray}
\tilde{Q}_{ij} = 6c/b Q_{ij}, \\ \nonumber \tilde{f}_b =
24^2c^3/b^4 f_b.
\end{eqnarray}
It will be understood that such scaling has been carried out, and we
shall omit the over-bars in the text below.
 We also introduce a
dimensionless temperature $\tau$ by defining
\begin{equation}
a(T-T^*) = \tau b^2/24c.
\end{equation}
The bulk nematic phase is stable 
for $\tau < \tau_{\rm NI} = 1$ with a degree of orientational order 
given by ${Q}_{b} = 3 (1+\sqrt{1-8\tau/9})/4$; ${Q}_b(\tau>1)=0$.
We model the temperature gradient in the $x$ direction by assuming
that $\tau$  changes linearly with the $x$
coordinate
\begin{equation}
  \tau = \tau_{NI} \left( 1 + \alpha x  \right).
\label{eq:grad}
\end{equation}
Equation~(\ref{eq:grad}) implies that the NI transition occurs at $x =
0$, and that the unconfined interface would form in the $yz$ plane.

We assume strong planar (parallel) anchoring of the director with the
particle surface. This is valid if the anchoring parameter $WR/K >>
1$, where $W$ is the anchoring energy of the particle surface, and
holds for relatively large colloidal particles or big anchoring
strengths. We also assume that the nematic phase at the particle
surface is uniaxial with a scalar order parameter $Q_{\rm s} =1$.

The elastic free energy density can be written
as~\cite{stephen.mj:1974.a}
\begin{equation}
f_e = \frac{1}{2} L_1 \frac{\partial Q_{ij}}{\partial x_k}
 \frac{\partial Q_{ij}}{\partial x_k} + \frac{1}{2} L_2
 \frac{\partial Q_{ij}}{\partial x_j} \frac{\partial Q_{ik}}{\partial x_k},
\label{eq:f_e}
\end{equation}
where the constants $L_1$ and $L_2$ are related to Frank-Oseen elastic
constants by $K_{11} = K_{33} = 9Q_b^2(L_1 + L_2/2)/2$ and $K_{22} =
9Q_b^2L_1/2$ and $Q_b$ is the bulk nematic order parameter.

We use typical values of the material parameters available for 5CB~\cite{coles.hj:1978,kralj.s:1991}:
$a = 0.044 \times 10^6 {\rm J/m^3K}$,
$b = 0.816 \times 10^6 {\rm J/m^3}$,
$c = 0.45 \times 10^6 {\rm J/m^3}$,
$L_1 = 6 \times 10^{-12} {\rm J/m}$, $L_2 = 2 L_1$,
$T^* = 307 {\rm K}$. The nematic-isotropic transition temperature 
for 5CB is $T_{NI} =308.5 \,{\rm K}$.

\subsection{Shape of the  interface}

%To account for the director deformations and anchoring (both at the
%interface and cell surfaces), the finite width of the interface, as
%well as the presence of defects,
 We minimize the free
energy~(\ref{eq:nem_free_en}) numerically, using finite elements with
adaptive meshes. The area $L_x \times  L_y$ is triangulated and the
order tensor $\bm Q$ is set at all vertices of the mesh and is 
linearly interpolated within each triangle~\cite{tasinkevych.m:2002.a,patricio.p:2002.a,andrienko.d:2004a}.
The free energy is then minimized using conjugate gradients method
under the constraints imposed by the boundary conditions. 

In the experiment, the $40\, \mu \rm m$ cells are filled with the
$16\, \mu \rm m$ colloidal particles. To minimize computational
efforts, we reduce the system size to $L_y = 3\, \mu \rm m$.  In
fact, minimization of the free energy in a system of this size is a
nontrivial task: the smallest length scale involved in the problem is
given by the thickness of the interface, which is of the order of the nematic
coherence length  at coexistence, 
$\xi = \left( 24 L_1 c/b^2 \right)^{1/2} \approx 10\, \rm nm$.  This
length scale sets the mesh size. Thus, about $10^7$ mesh points would
be required for a rectangular uniform discretization of the cell. By using adaptive meshes we are
able to reduce this number by more than two orders of magnitude, which
makes the minimization procedure computationally feasible.

\begin{figure}
\begin{center}
\includegraphics[width=8cm]{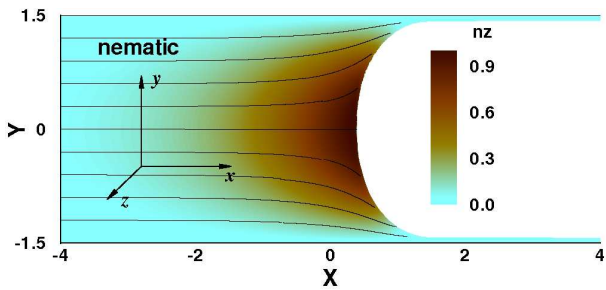}
\end{center}
\caption{Cross-section of the cell with the nematic-isotropic interface.
  Color coding shows the $n_z$, out of plane, component of the director.
  The director is practically parallel to the $z$ axis close to the
  interface (twist angle $\approx \pi / 2$, $n_z \approx 1$).
  $L_x = 15\mu {\rm m}, L_y = 3\mu \rm m$.   The temperature gradient
  parameter $\alpha=0.0036 \mu \rm m^{-1}$, which corresponds to $\approx 54 {\rm deg}/{\rm cm}$.}
\label{fig:numerics}
\end{figure}

We first have a look at the shape of the interface without colloidal
particle.  The chosen anisotropy of the elastic constants, 
$L_2/L_1 = 2$, favors  director alignment parallel to the NI 
interface, which is in mismatch with the boundary conditions at the 
cell boundaries. As we have already mentioned in the previous section,
this leads to formation of the curved  interface.
The contour plot of the $n_z$ (out of plane) component of the director
and the stream-traces of the in-plane ($n_x$ and $n_y$) components are shown in
Fig.~\ref{fig:numerics}. The interface is quasi-flat in the middle of 
the cell and curved at the horizontal surfaces of the box.
To avoid strong in-plane splay-bend deformations, the
director twists at the interface, in a qualitative agreement with the
experimental results. The twist angle at the interface
reaches $\pi/2$, i.~e. the director is parallel to the interface and the substrates.
Experimentally observed twist angle is smaller, of the order of
$\pi/4$. The difference between the experimentally measured and
calculated values is due to the different anchoring of the director at
the interface: for 5CB this angle is about $26\, \rm deg$, whereas our model
(which takes into account only second-order terms in the gradient
expansion) predicts planar interfacial anchoring.
The shape of the interface is symmetric, with its vertex
point placed in the middle between the walls, which is in  disagreement with the
experiment, where slightly asymmetric interfacial profile is observed.
 The asymmetry of the experimental interfaces 
is most probably due to the non-zero temperature gradient
across the cell (along the $y$ axis).

\subsection{Forces on a particle in the interfacial region}

\begin{figure}
\begin{center}
\includegraphics[width=7.5cm]{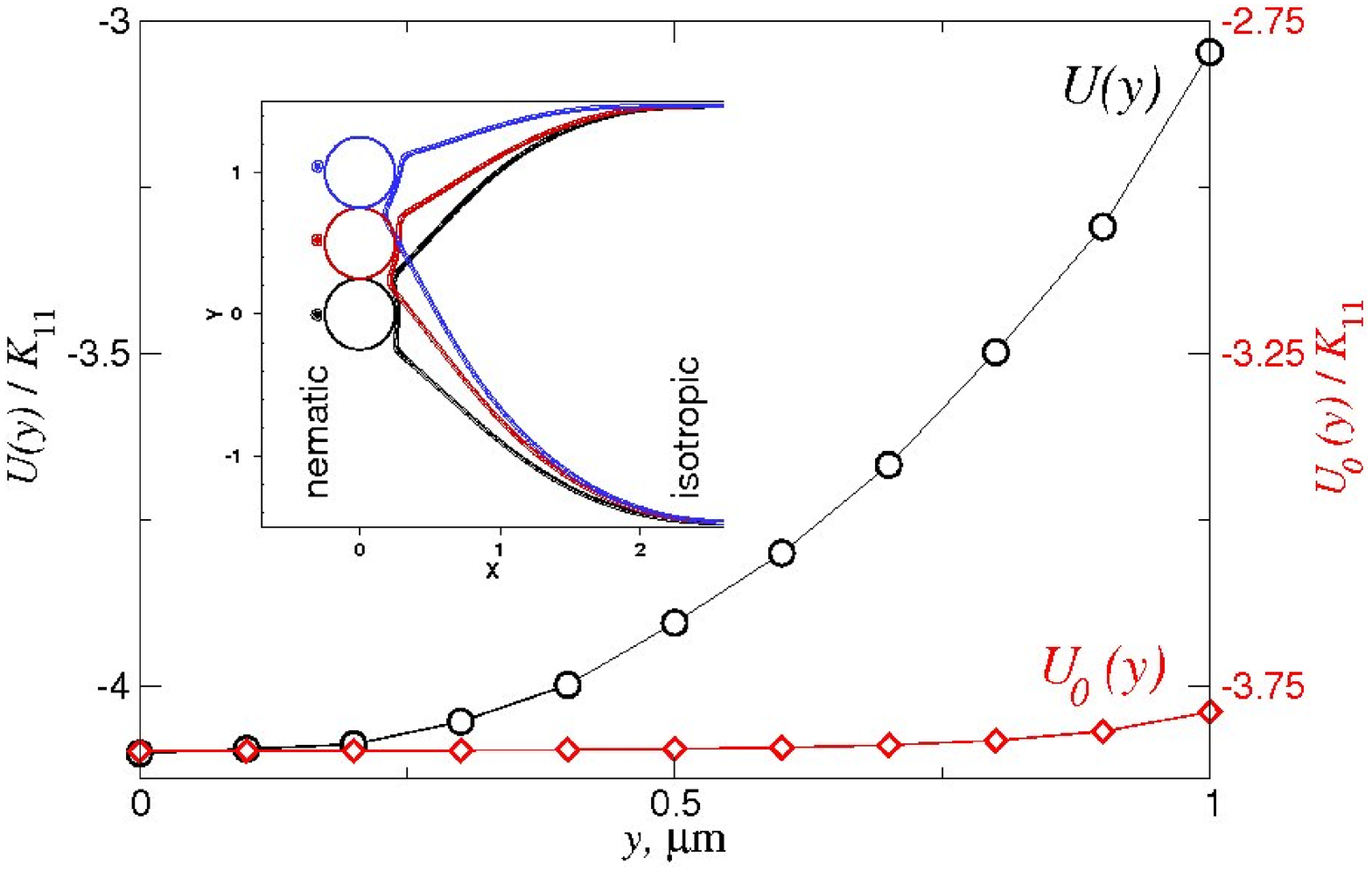}
\vspace*{0.3cm}

\hspace*{-0.95cm}\includegraphics[width=6.9cm]{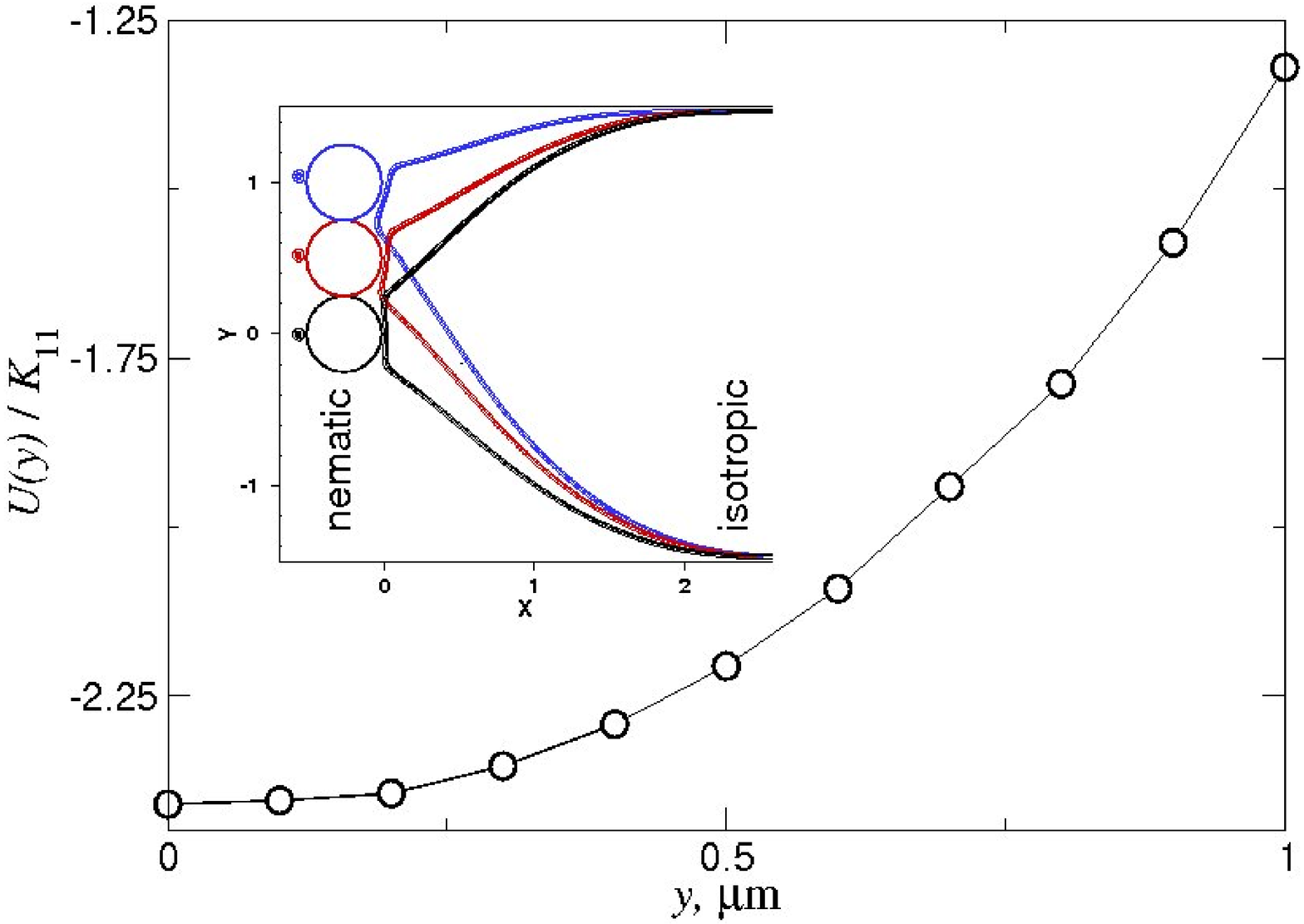}
\end{center}
\caption{Free energy $U(y)$ per unit length (circles). Colloid radius
  $R = 0.25\, \mu \rm m$, system size $9 \times 3\, \mu \rm m$.
  Temperature gradient $\alpha=0.001\mu \rm m^{-1}$ which corresponds
  to $\approx 15 {\rm deg}/{\rm cm}$.  Insets illustrate the shape of
  the interface for different positions of the colloidal particle.
  Upper panel: $x=0$; lower panel $x=-0.27 \mu \rm m$.  $U_0(y)$
  (shown in the upper panel, diamonds) is the free energy of a
  colloidal particle in a pure nematic phase, at $\tau = \tau_{NI}$.
  $K_{11} \approx 4.93 \times 10^{-12}\, \rm J/m$ for
  $\tau=\tau_{NI}=1$. }
\label{fig:energy}
\end{figure}

Now let us focus on the interaction of a colloidal particle with the NI
interface. In our previous work~\cite{west.jl:2002,andrienko.d:2004a}
we have shown that a single colloidal particle is attracted by a
{\em  flat} interface, with the force which is roughly
proportional to the particle radius. This force drags
the particles along the direction of the temperature gradient
($x$-axis). However, because of the wedge-like shape of the interface,
one can also expect an additional component of the force, along the $y$ axis.
To calculate this component, we placed a test particle, modeled as a long
cylinder of radius $R = 0.25\, \mu \rm m$ with the symmetry axis
parallel to the $z$ axis, next to the interface and evaluated the free
energy of the system as a function of the $y$-position of the
particle, at the fixed $x$-position. 
%We use rigid planar boundary conditions at the 
%colloidal surface.
 To improve the computational efficiency, the system size was
further decreased to $9 \times 3\, \mu \rm m$.

The free energy $U(x=const,y)\equiv U(y)$ per length of the colloidal
particle is shown in the Fig.~\ref{fig:energy} for two values of the
$x$-position of the colloid. The temperature gradient parameter
$\alpha=0.001\mu \rm m^{-1} $, corresponding to $\approx 15 {\rm
  deg}/{\rm cm}$. In both cases $U(y)$ has a minimum when the particle
is in the middle of the cell, at $y=0$. At any other point there is a
non-zero force $ F_y = - \partial {\cal F} / \partial y$, which pulls
the particle towards the middle of the cell (upwards, if it is placed
at the bottom wall).  Therefore, even when the interface is not
moving, those particles which are captured by the interface tend to
cluster in the middle of the cell, at the vertex line.
%Moreover, if the
%interface moves, particles could experience an additional force directed
%towards the vertex line, due to the capillary force which is normal to
%the curved interface and directed inward the growing phase. 
 
Because of the fixed anchoring conditions imposed on both the cell and
colloid boundaries, one might expect that the particle is repelled
from the cell walls even without the presence of a NI interface. To
check this, we have also calculated the free energy $U_0(y)$ of a
colloidal particle immersed in a pure nematic phase ($\tau=\tau_{NI}$,
$\alpha=0$, same cell size and boundary conditions as before). Our
calculations (see the upper panel of Fig.~\ref{fig:energy})
demonstrate that this force is by the order of magnitude smaller,
i.~e. the presence of the curved interface plays a dominant role.

\section{Conclusions}
To summarize, our experiments demonstrate that the curved shape of the
interface and the nonuniform director distribution next to it result
in a reach dynamics of colloidal particles in the interfacial region.
We show that the particles can be dragged by the interface not only
horizontally, across the cell, but can also be moved in the vertical
direction.  This effect can be used to build heterogeneous
three-dimensional photonic crystals, where the impurities (or holes)
shall be positioned in designed places~\cite{west.jl:2004.a}.

The drag effect can also be used to design composite materials with
complex morphologies. The moving interface is able to rearrange the
nucleation centers of, for example, phase-separating polymer network.
During polymerization, spatial distribution of these centers affects
the final morphology of the network.  This opens up an opportunity to
create composite materials of extraordinary properties.

\acknowledgments

We thank to V.~Reshetnyak and A.~Iljin for useful discussions.  The
work was partially supported by the DARPA grant No. 444226 and by the
grant ``Composite liquid crystal and polymer materials for information
technologies'' of National Academy of Science of Ukraine.

\bibliography{paper_2col}

\end{document}